\documentclass[useAMS,usenatbib,longnamesfirst,usedcolumn]{mn2e} 
\usepackage{epsfig}	
\usepackage{times}	
\usepackage{amsmath,epsf}
\title[{\it Chandra} and {\it XMM-Newton} Observations of Abell 3158]{A Joint {\it Chandra} and {\it XMM-Newton} View of Abell 3158: Massive Off-Centre Cool Gas Clump As A Robust Diagnostic of Merger Stage}
\author[Yu Wang et al.]{Yu Wang$^{1}$, 
Haiguang Xu$^{1}$, 
Liyi Gu$^{1}$, 
Junhua Gu$^{1}$, 
Zhenzhen Qin$^{1}$, 
\newauthor Jingying Wang$^{1}$, 
Zhongli Zhang$^{2}$
and Xiang-Ping Wu$^{3}$\\
$^{1}$Department of Physics, 
Shanghai Jiao Tong University, 
800 Dongchuan Road, 
Shanghai 200240, 
PRC. E-mail: wenyu$_{-}$wang@sjtu.edu.cn; hgxu@sjtu.edu.cn\\
$^{2}$Max-Planck-Institute of Astrophysics, 
Karl-Schwarzschild-Str. 1, 
Postfach 1317 D-85741, 
Garching, Germany\\
$^{3}$National Astronomical Observatories, 
Chinese Academy of Sciences, 
20A Datun Road, 
Beijing 100012, 
PRC\\
\\}
\date{Received October 22 / Accepted 30 December 2009}
\pagerange{\pageref{firstpage}--\pageref{lastpage}}
\pubyear{2009}

\begin{document}

\maketitle

\label{firstpage}

\begin{abstract}
\noindent By analysing the {\it Chandra} and {\it XMM-Newton} archived 
data of the nearby galaxy cluster Abell 3158, which was reported to 
possess a relatively regular, relaxed morphology in the X-ray band 
in previous works, we identify a bow edge-shaped discontinuity in the 
X-ray surface brightness distribution at about $120h_{71}^{-1}$ kpc west 
of the X-ray peak. This feature is found to be associated with a 
massive, off-centre cool gas clump, and actually forms the west boundary 
of the cool clump. By calculating the thermal gas pressures in the cool 
clump and in the free-stream region, we determine that the cool gas clump 
is moving at a subsonic velocity of 
$700^{+140}_{-340}$ km s$^{-1}$ ($\mathcal M=0.6^{+0.1}_{-0.3}$) 
toward west on the sky plane. We exclude the possibility that this cool 
clump was formed by local inhomogeneous radiative cooling in the intra-cluster 
medium, due to the effectiveness of the thermal conduction on the time-scale 
of $\sim 0.3$ Gyr. Since no evidence for central AGN activity has been found 
in Abell 3158, and this cool clump bears many similarities to the off-centre 
cool gas clumps detected in other merging clusters in terms of their mass, 
size, location, and thermal properties (e.g. lower temperature and higher 
abundance as compared with the environment), we speculate that the cool clump 
in Abell 3158 was caused by a merger event, and is the remnant of the original 
central cool-core of the main cluster or the infalling sub-cluster. This idea 
is supported not only by the study of line-of-sight velocity distribution of 
the cluster member galaxies, but also by the study of gas entropy-temperature 
correlation. This example shows that the appearance of such massive, off-centre 
cool gas clumps can be used to diagnose the dynamical state of a cluster, 
especially when prominent shocks and cold fronts are absent.
\end{abstract}


\section{INTRODUCTION}
In the frame of hierarchical clustering scenario, galaxy clusters 
grow in size by merging with subunits, with each major merger 
event lasting for $2-5$ Gyr (e.g. Roettiger, Loken \& Burns 1997; Ascasibar 
\& Markevitch 2006; Poole et al. 2006). In such major mergers both 
ram pressure stripping and slingshot can generate remarkable X-ray 
substructures in gas density and temperature distributions, which 
are often accompanied by shocks and/or cold fronts that exhibit 
arc-shaped or edge-like morphologies (e.g. Markevitch \& Vikhlinin 
2007 and references therein). These substructures are expected to 
contain valuable information not only on merger dynamics itself, 
but also on thermal and chemical evolutions of the hot intra-cluster 
medium (ICM), which helps understand the nature of dark matter 
(Markevitch et al. 2004; Randall et al. 2008).

Theoretical studies and numerical N-body simulations (e.g.
Roettiger et al. 1997; Ricker 1998; Poole et al. 2006) showed 
that in a typical major merger event temperature substructures 
may survive for a relatively long time of 
$t_{\rm thermal}\simeq n_{\rm gas}kT/n_{\rm gas}^{2}\Lambda$,
where $n_{\rm gas}$, $T$, and $\Lambda$ are gas density, temperature,
and cooling function, respectively.
The derived $t_{\rm thermal}$ (up to about 4 Gyr in some cases) 
is much longer than the 
lifetimes of shocks and cold fronts, since shocks and cold fronts 
are usually smeared out quickly on dynamic time-scales characterized 
by the sound crossing time. Actually, high-quality {\it Chandra} 
and {\it XMM-Newton} observations have shown that in merging 
clusters there is a prevailing existence of massive, off-centre 
cool gas clumps, which are about $50-300$ kpc in size, reside 
within the central ($\la$600 kpc) region of the cluster, 
and possess an average temperature $1-4$ keV lower than the 
ambient gas.

In 1E $0657-56$ (the Bullet cluster; Markevitch et al. 2002), Abell 
520 (Govoni et al. 2004), Abell 2256 (Sun et al. 2002), and Abell 
3667 (Vikhlinin, Markevitch \& Murray 2001), the off-centre cool gas clump is found 
to be moving adiabatically behind a bow shock and/or a cold front, 
and thus can be directly identified as the remnant of the original
central cool-core of the infalling sub-cluster. On the other hand, 
in clusters such as Abell 754 (Markevitch et al. 2003), Abell 2065 
(Chatzikos, Sarazin \& Kempner 2006), and Abell 2255 (Sakelliou \& Ponman 2006), 
the off-centre cool gas clump is not escorted by either a shock or 
a cold front. However, unambiguous evidence obtained in radio, 
optical, and X-ray bands still indicates that the cool gas clump 
was driven out of the core region of either the main cluster (Abell 
754; Markevitch et al. 2003; Henry, Finoguenov \& Briel 2004) or the infalling 
sub-cluster by the merger.

In fact, the off-centre cool gas clumps in above two types of clusters
share many similarities in terms of their sizes, locations, and gas
masses ($\sim10^{11}$ ${\rm M}_{\odot}$). Therefore, an interesting question
arises: can the appearance of such massive, off-centre cool gas clumps 
be safely used to diagnose the dynamical state of a cluster, even when 
there is a lack of violent merger signatures, such as prominent shocks,  
cold fronts, and other associated radio/optical substructures? Clearly, 
in order to address this issue, we need to examine gas temperature 
distribution in more galaxy clusters that show only weak merger signatures 
and have evolved into late merger stages (usually mean post core passages).

In this work we present a detailed joint {\it Chandra} and {\it 
XMM-Newton} study of the nearby rich galaxy cluster Abell 3158 
(richness class 2, Quintana \& Halven 1979; $z=0.0597$, Struble \& 
Rood 1999), which contains 105 member galaxies within the central 
$2.4h_{71}^{-1}$ Mpc (Biviano et al. 2002). In literature (e.g. 
Irwin, Bregman \& Evrard 1999; Lokas et al. 2006; Chen et al. 2007), this cluster 
is usually referred as a relaxed, non-cool-core system, because 
neither significant X-ray substructures nor diffuse radio emission 
has been reported in the previous works 
(Ku et al. 1983; Mohr, Mathiesen \& Evrard 1999; 
Schuecker et al. 2001; Mauch et al. 2003). In the optical band, 
however, the cluster shows some unusual properties that can be possibly 
related to a recent merger event, as are summarized as follows.
First, the cluster hosts three luminous elliptical galaxies that are
brighter than $9\times10^{10}$ ${\rm L_{B,\odot}}$ (Paturel et al. 2003), 
all classified as a cD galaxy by S$\acute{\rm e}$rsic (1974). Two of them, the brightest 
member PGC 13641 
(E0, $L_{B}=1.45\times10^{11}$ ${\rm L_{B,\odot}}$) 
and the S0 galaxy PGC 13652 
($L_{B}=9.39\times10^{10}$ ${\rm L_{B,\odot}}$), 
constitute a closely aligned central galaxy pair, whose projected 
separation is $81.3h_{71}^{-1}$ kpc, and jointly dominate the cluster 
(Quintana \& Havlen 1979). The second luminous member PGC 13679 
(S0, $L_{B}=9.48\times10^{10}$ ${\rm L_{B,\odot}}$) 
is located at about 417.8$h_{71}^{-1}$ kpc southeast of the central
galaxy pair, and dominates a local sub-cluster whose recessional 
velocity is about 1 300 km s$^{-1}$ larger than the mean recessional 
velocity of the main cluster (17 500 km s$^{-1}$; Lucey et al. 1983; 
Kolokotronis et al. 2001).
Second, as shown in figure 7 of Smith et al. (2004), the line-of-sight 
velocity distribution of the cluster's member galaxies exhibits a
plateau at the high-velocity end, which is often seen in merging systems.
Third, Kolokotronis et al. (2001) reported that in this cluster there 
is an offset of about $200h_{71}^{-1}$ kpc between the optical 
flux-weighted centroid and the optical density peak. 
Based on these we speculate that Abell 3158 is evolving at a late merger 
stage and is thus an ideal target for our investigation.

Throughout the paper, we adopt the cosmological parameters 
$H_{0} = 71$ km s$^{-1}$ Mpc$^{-1}$, 
$\Omega_{\rm M} = 0.27$, and
$\Omega_{\Lambda} = 0.73$, 
so that $1^{\prime\prime}$ corresponds to about $1.14 h_{71}^{-1}$ kpc 
at the redshift of the cluster. We utilize the solar abundance standards 
given by Grevesse and Sauval (1998), where the iron abundance relative 
to hydrogen is $3.16\times 10^{-5}$ in number. Unless stated otherwise, 
the quoted errors are the 90$\%$ confidence limits.

\section{OBSERVATIONS AND DATA REDUCTIONS}
\subsection{\it Chandra}
Except for an extremely short exposure on 2007 September 16
(5.1 ks, ObsID 7688), which is not used in this work, Abell
3158 has also been observed with {\it  Chandra} on 2002 June 19 
(31.4 ks, ObsID 3712) and June 21 (25.1 ks, ObsID 3201),  
respectively, with chips 0, 1, 2, 3, and 6 of the Advanced CCD 
Imaging Spectrometer (ACIS) operating in VFAINT mode. We used 
the {\it Chandra} data analysis package \textsc{ciao} version 4.1 and 
\textsc{caldb} version 4.1.2 to process the archived data in the standard 
way, by starting with the level-1 event files. Corrections for 
the charge transfer inefficiency and time dependent gain have 
been applied. We kept events with {\it ASCA} grades 0, 2, 3, 4, 
and 6, and removed all the bad pixels, bad columns, and columns 
adjacent to bad columns and node boundaries. We examined the 
$0.3-10.0$ keV lightcurves extracted from the background regions 
defined on ACIS chips 0 and 1, and found that there are no strong 
flares that increase the background count rate to $>120\%$ of 
the mean quiescent value. The obtained net exposures are 30.9 
ks and 24.8 ks for the two observations, respectively. In the 
spectral analysis, we extracted the {\it Chandra} spectra in 
$0.7-8.0$ keV from the two observations separately, and fitted 
them simultaneously with the same model, except that their 
normalizations were left free. Background spectra were extracted 
from the {\it Chandra} blank-sky fields; a crosscheck based on 
the use of local background yielded essentially the same results.

\subsection{\it XMM-Newton}
Abell 3158 was observed with {\it XMM-Newton} on 2005 November 22
(22.4 ks, ObsID 0300210201) and 2006 January 18 (9.4 ks, ObsID
0300211301), respectively, with the European Photon Imaging Camera 
(EPIC) operating in PrimeFullWindow (MOS1 and MOS2) and PrimeFullWindowExtended
(pn) modes. Although during the second observation the Reflection 
Grating Spectrometer (RGS) was also turned on, in this work we 
limited our analysis to the first observation only by using the latest 
version of \textsc{sas} (version 8.0.1) and its standard filterings. We  
kept the events with \textsc{flag}=0 and \textsc{pattern}s $0-12$ for MOS1 and 
MOS2, and the events with \textsc{flag}=0 and \textsc{pattern}s $0-4$ for pn. We 
removed all the time intervals contaminated by soft proton flares,
during which the $10-12$ keV count rate exceeds the $2\sigma$ 
limit of the mean quiescent value (see, e.g. Katayama et al. 
2004). The cleaned MOS1, MOS2, and pn datasets have effective 
exposure times of 20.9 ks, 20.6 ks, and 12.2 ks, respectively. 
In the spectral analysis, we removed all the bright X-ray point 
sources and adopted conservative energy cuts at 0.5 keV and 
8.0 keV. The obtained MOS1, MOS2, and pn spectra were fitted 
simultaneously using the same model, except that their 
normalizations were left free. Backgrounds spectra were generated 
from the blank-sky event lists, which were filtered in advance 
using the same selection criteria (\textsc{pattern}, \textsc{flag}, etc.) as used 
for the source spectra.

\section{X-RAY ANALYSIS AND RESULTS}
\subsection{X-ray surface brightness discontinuity}
In Fig. $1a$ and $1b$, we show the {\it Chandra} ACIS and combined 
{\it XMM-Newton} EPIC images of the central 
$1.5h_{71}^{-1}$ Mpc$\times 1.5h_{71}^{-1}$ Mpc ($21^{\prime}.9\times21^{\prime}.9$) 
region of Abell 3158 in $0.3-2.0$ keV, respectively, which have been 
adaptively smoothed and corrected for both exposure and vignetting. 
The {\it XMM-Newton} image was generated by combining the MOS1, MOS2, 
and PN events together using the \textsc{sas} `image' script.
The X-ray peak of the cluster 
(R.A.=03h42m52.7s Dec=$-$53d37m37.2s J2000) 
is found at about 
$17.8h_{71}^{-1}$ kpc ($0^{\prime}.26$) 
northwest of the optical centroid of PGC 13641 
(R.A.= 03h42m52.9s Dec=$-$53d37m52.2s; Katgert et al. 1998), 
the brighter member of the central galaxy pair. The corresponding 
optical Digital Sky Survey (DSS) image is shown in Fig. $1c$, on 
which the {\it Chandra} intensity contours obtained from Fig. $1a$
are overlaid.

On both {\it Chandra} and {\it XMM-Newton} images it can be clearly 
seen that the X-ray isophotes are elongated in nearly the east-west 
direction with an ellipticity of $\simeq0.3$. And on the {\it Chandra} 
image a bow edge can be identified at about 
$120h_{71}^{-1}$ kpc ($1^{\prime}.88$) 
west of the X-ray peak. According to its shape and location on the 
detector, the possibility of the feature being associated with the
artifacts caused by CCD gaps or bad columns can be excluded.

In order to examine the significance of the bow edge in a quantitative
way, we define two sets of partial elliptical annuli, which are confined 
in the east and west sector regions shown in Fig. $2a$, and extract the 
exposure-corrected X-ray surface brightness profiles (SBPs) with both
{\it Chandra} ACIS and {\it XMM-Newton} EPIC (Fig. $2b$). Because the 
gas temperature of Abell 3158 is sufficiently high ($5.8$ keV; Reiprich 
\& B$\ddot{\rm o}$hringer 2002), by following, e.g. Markevitch et al. 
(2000), we restricted the SBP extractions in $0.3-2.0$ keV to minimize 
the dependence of X-ray emissivity on possible temperature fluctuations 
(see \S3.2). We attempt to apply the standard $\beta$ model to describe
the extracted SBPs, which is expressed as 
\begin{equation}
S(r) =S_{0}[1+(r/r_{\rm c})^{2}]^{0.5-3\beta}+S_{\rm bkg}\\
\end{equation}
(Jones \& Forman 1984), 
where $r_{\rm c}$ is the core radius, $\beta$ is the slope, 
$S_{\rm bkg}$ is the background, and $S_{0}$ is the normalization.
We find that in the east sector the SBPs can be
fitted very well with the $\beta$ model (Table 1), whereas in the west 
sector the SBPs show a significant central excess beyond the best-fit 
$\beta$ model for the outer ($>120h_{71}^{-1}$ kpc) regions. 
Within about $90h_{71}^{-1}-115h_{71}^{-1}$ kpc 
($1^{\prime}.32-1^{\prime}.68$), where 
the bow edge is identified visually, there exists a sudden discontinuity, 
across which the {\it Chandra} and {\it XMM-Newton} surface brightness 
increases inwards by a factor of $\simeq1.6$ and $\simeq1.3$, respectively; 
this difference may be caused by the different distributions of the point
spread function (PSF) of {\it Chandra} ACIS and {\it XMM-Newton} EPIC. 
These results confirm the existence of the bow edge shown in Fig. $1a$, 
and indicate that it is not related to the central surface brightness 
excess usually seen in the cD galaxies (Makishima et al. 2001), which 
are not associated with a surface brightness discontinuity. Also, we
note that the X-ray surface brightness discontinuity in Abell 3158 is 
similar to, although not as sharp as, those found in 1E $0657-56$ and 
Abell 3667, in which the jumps of the surface brightness at the edges 
are $2-3$ within $\simeq10h_{71}^{-1}$ kpc and thus the edges have been 
confirmed as cold fronts caused by mergers 
(Markevitch et al. 2002; Vikhlinin et al. 2001).

\subsection{An unusual off-centre cool gas clump}
In order to investigate the thermal properties of the bow edge, we 
calculate the two dimensional gas temperature distribution in the 
central 
$0.6h_{71}^{-1}$ Mpc$\times 0.6h_{71}^{-1}$ Mpc 
($8^{\prime}.77\times8^{\prime}.77$) 
region of Abell 3158, by following the approach of, e.g. O'Sullivan 
et al. (2005) and Gu et al. (2009). In the calculation only the {\it 
XMM-Newton} EPIC datasets are used, because the data statistics of 
the {\it Chandra} ACIS datasets is not good enough for us to map the 
temperature distribution. To be specific, we first define about 4000 
cells in the 
$0.6h_{71}^{-1}$ Mpc$\times 0.6h_{71}^{-1}$ Mpc
region, whose centres (${\bf r}_{i}$, $i=1, 2, 3 ...$) are randomly 
distributed with a separation of $<10^{\prime\prime}$ between the 
centres of any two adjacent cells. Each cell is apportioned with an 
adaptive radius of $0^{\prime}.3-0^{\prime}.8$, so that it encloses 
$>3000$ photons in $0.5-8.0$ keV after the point sources are excluded. 
Then we extract the MOS1, MOS2, and pn spectra from each cell and fit 
them simultaneously with an absorbed \textsc{apec} model coded in \textsc{xspec} v12.4.0. 
The redshift and absorption are fixed to $z=0.0597$ and the Galactic 
value $N_{\rm H} = 1.62 \times 10^{20}$ cm$^{-2}$ (Dickey \& Lockman 
1990), respectively. After the obtained best-fit gas temperature is 
assigned to the corresponding cell ($T_{\rm c}({\bf r}_{i})$), and 
this process is repeated for all cells, we calculate the projected 
temperature at any position ${\bf r}$ as 
\begin{equation}
T({\bf r})=\sum_{{\bf r}_{i}}{[G_{{\bf r}_{i}}(R_{{\bf r,r}_{i}})T_{\rm c}({\bf r}_{i})]}/\sum_{{\bf r}_{i}}{G_{{\bf r}_{i}}(R_{{\bf r,r}_{i}})},~R_{{\bf r,r}_{i}}<s({\bf r}_{i})\\
\end{equation}
(Gu et al. 2009), where $R_{{\bf r,r}_{i}}$ is the distance between 
${\bf r}$ and the centre of the cell $i$, $s({\bf r}_{i})$ is 
defined as the radius of cell $i$, and $G_{{\bf r}_{i}}$ is the Gaussian 
kernel with a scale parameter $\sigma=s({\bf r}_{i})$.
The obtained temperature map, along
with the $1\sigma$ error map that is calculated in nearly the same 
way as the temperature map, are shown in Fig. $3a$ and $3b$, 
respectively.

We find that, despite the relatively regular appearance of the X-ray
images (Fig. 1), the spectral distribution of the projected gas 
temperature is highly asymmetric, inferring that the cluster has 
not recovered from a violent event on cluster scales. In particular, 
we find that there exists a cool gas clump at about 
$84h_{71}^{-1}$ kpc ($1^{\prime}.23$) 
west of the X-ray peak, whose linear scale is about
$80h_{71}^{-1}$ kpc ($1^{\prime}.17$). 
The projected gas temperature within this cool clump ranges from 
about 4.4 keV to 5.0 keV, which is apparently lower than that of the 
ambient gas ($5.2-6.0$ keV). Most interestingly, the west boundary 
of the cool gas clump coincides with the bow edge perfectly, indicating
that they are likely to have the same origin.

To clarify the significance of the existence of the cool gas clump,
we define four sector regions in Fig. $3a$, two of which have the same angles as
those defined earlier in Fig. $2a$, and divide each of these sector regions
into a series of wider partial elliptical annuli. We extract both the {\it 
Chandra} ACIS and {\it XMM-Newton} MOS1, MOS2, and pn spectra 
from these annuli, and fit them with an absorbed \textsc{apec} model when 
the redshift and absorption are fixed again as above; allowing the 
absorption to vary does not improve the fits. In the fittings, the
{\it Chandra} ACIS spectra extracted from the two observations are treated
as group one, and the {\it XMM-Newton} MOS1, MOS2 and pn spectra are treated
as group two. Each group of spectra are fitted simultaneously with the 
same model parameters, except for that the normalizations are left free.
The derived projected temperature profiles in the four sectors 
are plotted in Fig. $4a-4d$. It can be clearly seen that the {\it 
Chandra} and {\it XMM-Newton} temperatures are consistent with each 
other (68\% confidence level), and in the west sector the temperature 
in 
$45.4-95.3h_{71}^{-1}$ kpc ($4.85\pm0.17$ keV), 
where the cool gas clump is located, is lower than those of the 
adjacent regions at 68\% confidence level (see also Table 2); 
no such temperature drop is found in other three directions. 
This is further confirmed by the results of the deprojected analysis 
of the {\it XMM-Newton} spectra for the west sector (Table 2 and Fig. $4e-4f$), 
which shows that the gas temperature of the cool gas clump 
($3.98^{+0.43}_{-0.41}$ keV for W2 region) 
is actually lower than those of the innermost region 
($5.71^{+1.04}_{-0.71}$ keV for W1 region) 
and the next outer region 
($5.38^{+0.74}_{-0.51}$ keV for W3 region) 
at the $90\%$ confidence level, and cannot be ascribed to the 
uncertainties in determining the metal abundance.

\subsection{Faint Cold front associated with the cool gas clump}
To study the nature of the cool gas clump, we carry out a simple 
hydrodynamic study to determine the velocity of the edge in the
environment. By applying the best-fit deprojected spectral 
parameters (Table 2), here we revisit the X-ray surface brightness 
profiles of the west sector (Fig. $2b$) and fit it again with 
two models. In the first model (model A), the spatial distribution
of gas density is described with two $\beta$ components as
\begin{equation}
n_{\rm gas}(R) =\{ n^{2}_{\rm g,1}[1+(R/R_{\rm c1})^{2}] ^{-3\beta_{1}} 
+ n^{2}_{\rm g,2}[ 1 + (R/R_{\rm c2})^{2} ]  
^{-3\beta_{2}}\}^{1/2},
\end{equation}
where $R$ is the 3-dimensional radius, $R_{\rm c}$ is the core 
radius, and $\beta$ is the slope. In the second model (model B), 
the gas density distribution is broken at the truncation radius 
$R_{\rm cut}$, described with one truncated power-law component 
and one truncated $\beta$ component as
\begin{equation}
   n_{\rm gas}(R) = \left \{ \begin{array}{ll}
       n_{\rm g,1}(R/R_{\rm cut})^{-\alpha} & R<R_{\rm cut} \\
       n_{\rm g,2} [ 1 + (R/R_{\rm c})^{2} ] ^{-3\beta/2}& R  
\geq R_{\rm cut}
                     \end{array}
                   \right.  .
\end{equation}
Acceptable fits to both the {\it Chandra} and {\it XMM-Newton} SBPs 
can be obtained with model B only (Table 1). The {\it Chandra} best-fit 
suggests a density jump by a factor of $f_{\rm jump}=1.7\pm0.1$ at 
$R_{\rm cut}=111.0\pm4.4h_{71}^{-1}$ kpc
(roughly the position of the edge), and the {\it XMM-Newton} best-fit  
gives consistent values of $f_{\rm jump}=1.5\pm0.1$ and 
$R_{\rm cut}=104.9\pm2.2h_{71}^{-1}$ kpc.

Following Vikhlinin et al. (2001), we use the best-fit gas densities 
across the edge (Eq.(4)) and gas 
temperatures for W2 and W3 regions (\S3.2 and Fig. 4$e$)
to calculate the thermal gas pressures in the cool gas clump ($P_{0}$) 
and in the free-stream region ($P_{1}$), respectively. 
The former pressure is assumed to be in pressure equilibrium
with the gas pressure at the stagnation point (denoted as the place where
the relative gas velocity vanishes). 
The pressure ratio is estimated to be $P_{0}/P_{1}=1.3\pm0.2$ (68\% 
confidence level). This allows us to determine the Mach number ($\mathcal M$) 
of the cool clump to be $0.6^{+0.1}_{-0.3}$, which corresponds to a velocity
of $700^{+140}_{-340}$ km s$^{-1}$. 
This subsonic velocity of the clump, along with the results obtained in 
\S3.1 (i.e., surface brightness discontinuity) and \S3.2 (i.e., gas 
temperature and density discontinuities), allow us to conclude that 
there exists a faint cold front at the west boundary of the cool gas clump.

\section{VELOCITY PLATEAU}

In early works of Biviano et al. (1997 and 2002), Girardi et al. (1996), 
and Kolokotronis et al. (2001),
both the two dimensional spatial distribution and the  
line-of-sight velocity distribution of the member galaxies in 
Abell 3158 were studied, and no significant substructure was 
reported. However, after tighter and more reliable selection 
criterion for member galaxies, which was based on the galaxy 
colour-magnitude relation (e.g. Colless \& Dunn 1996; Ferrari 
et al. 2003; Maurogordato et al. 2008), was applied, Smith 
et al. (2004) re-constructed the line-of-sight velocity 
distribution with higher fidelity and found that there exists 
a visible plateau beyond the Gaussian distribution at the 
high-velocity side. Since Smith et al. (2004) did not use a 
quantitative approach to assess the significance of the 
high-velocity plateau, in which the merger information might 
be contained, here we draw the velocity data of the member 
galaxies in Abell 3158 from Smith et al. (2004), and re-analyse 
the line-of-sight velocity distribution profile. After the 
model fitting of this distribution is finished, we also attempt 
to investigate the 2-dimensional spatial distribution of the 
galaxies in the high-velocity plateau.

First, we plot the line-of-sight velocity distribution of the 68 
member galaxies identified by Smith et al. (2004) in Fig. $5a$, 
which shows that the high-velocity plateau is located at about 
$19~000$ km s$^{-1}$. We fit the observed distribution with a 
single Gaussian profile, and calculate the Kolmogorov$-$Smirnov  
statistic for the observed distribution against the best-fit Gaussian 
model ($\chi^{2}/dof=31.6/9$). We find that the observed distribution 
has a probability of $<10\%$ of being Gaussian. We then attempt to 
fit the observed distribution with a two-Gaussian model. The best-fit 
($\chi^{2}/dof=9.8/6$) gives an average velocity of 
$<v_{1}>=17~380\pm50$ km s$^{-1}$
and a corresponding variance of 
$\sigma_{v,1}=520\pm50$ km s$^{-1}$
for the main Gaussian component, and 
$<v_{2}>=19~120\pm130$ km s$^{-1}$
and 
$\sigma_{v,2}=510\pm100$ km s$^{-1}$
for the high-velocity plateau.
By applying the F-test and the Kaye's Mixture Model (KMM; McLachlan 
\& Basford 1988; Ashman, Bird \& Zepf 1994) test, the latter of which 
is based on a maximum likelihood algorithm, we find that the second 
Gaussian component is required at 94.3\% confidence level and preferred 
at a significant probability of 90\%, respectively.

To investigate if the galaxies in the high-velocity plateau form 
a real substructure in the cluster, we divide the member galaxies 
into two subgroups: one low-velocity ($15~300-18~500$ km s$^{-1}$) 
subgroup and one high-velocity ($18~500-20~000$ km s$^{-1}$) subgroup, 
which consist of 51 and 17 galaxies, respectively. According to the 
best-fit two-Gaussian model, these two subgroups roughly corresponds 
to the two Gaussian components, respectively, with up to about two
of the galaxies in the high-velocity subgroup coming from the main 
Gaussian component. We find that the galaxies belonging to the 
high-velocity subgroup are distributed mostly in the central 
$500h_{71}^{-1}$ kpc (7$^{\prime}$.3), as well as the west part 
of the cluster, with a geometric centroid of the high-velocity 
galaxies is located at about $355h_{71}^{-1}$ kpc (5$^{\prime}$.2) 
northwest of the X-ray peak, which is not coincident with any 
X-ray substructures or luminous member galaxies.
The high-velocity subgroup roughly includes the member galaxises  
of the PGC 13679 sub-cluster, since the member galaxies and the 
boundary of the latter are difficult to be determined by 2-dimensional 
spatial distribution alone (e.g. Colless \& Dunn 1996; Boschin et al. 2006).
The galaxies in the low-velocity subgroup that is dominated by the central 
PGC 13641-PGC 13652 galaxy pair, on the other hand, are 
scattered symmetrically in the field. These results suggest that 
the galaxy velocity separation (Fig. $5a$; see also Smith et al. 2004) 
has a dynamical nature, which is most likely to be a major merger, 
with a merger mass ratio of about $1:1-3$, as estimated from 
the velocity variance ratio ($\sigma_{v,2}/\sigma_{v,1}\simeq1:1$) 
and the galaxy number ratio ($\simeq 17/51=1:3$) of the two subgroups.
    
In literature Abell 3158 was classified as a `single-component'
and `relaxed' system (Kolokotronis et al. 2001; Lokas et al. 2006),
which possesses a relatively regular morphology in the X-ray band
(Ku et al. 1983; Mohr et al. 1999; Schuecker et al. 2001). 
However, considering none of violent merger signatures is detected in the cluster,
the existence of the high-velocity subgroup indicates that the cluster 
is evolving into the late stage of a major merger event. This may explain 
the origin of the off-centre cool gas clump detected in the west sector 
region, which is moving at a subsonic velocity (\S3.2$-$\S3.3; see also \S5.1).

\section{DISCUSSION}
\subsection{Merging origion of the massive, off-centre cool gas clump}

The most prominent X-ray feature detected in Abell 3158 is the 
massive, off-centre cool gas clump, which is moving behind a {\bf faint} cold front. 
It does not resemble the cool substructures caused by AGN activity 
(e.g. Forman et al. 2005; Nulsen et al. 2002), which are usually tightly
associated with the low-density regions such as the AGN-induced 
radio lobes and X-ray cavities, since no evidence for AGN activity 
and its remnants has been reported in Abell 3158 according to 
the existing multi-band observations. In such a sense, we speculate 
that the cool clump was formed either by the inhomogeneous radiative 
cooling in the ICM, or by the ram pressure stripping or slingshot 
during a merger event. The first possibility can be ruled out 
immediately, because given the best-fit spectral parameters for 
the cool clump, the radiative cooling time of the gas in the clump 
(cf. Sarazin 1986) is 
\begin{eqnarray}
t_{\rm cool} &\; =\; & 3.5\times10^{10}\;~{\rm yr}~ \times 
\left( \frac{n_{\rm gas}}{2\times 10^{-3}\; {\rm cm}^{-3}}\right)^{-1} 
\nonumber \\
             & \times & 
\left( \frac{kT}{10~{\rm keV}} \right)^{1/2}
~\simeq9.8~{\rm Gyr},
\end{eqnarray}
where continuum and line emissions are both included.
This is close to the Hubble time ($\simeq12.9$ Gyr) of the cluster, 
and is significantly longer than the Coulomb conduction time
\begin{eqnarray}
t_{\rm cond} &\; \sim\; & 9.2\times10^{6}\; ~{\rm yr}~ \times 
\left( \frac{n_{\rm gas}}{2\times 10^{-3}\; {\rm cm}^{-3}}\right)~\nonumber \\
   & \times & 
\left( \frac{l}{100~{\rm kpc}} \right)^2
\left( \frac{kT_{\rm centre}}{10~{\rm keV}} \right)^{-5/2}
\simeq0.06~{\rm Gyr}
\end{eqnarray}
(Markevitch et al. 2003), where $l=80h_{71}^{-1}$ kpc is the 
linear scale of the cool clump, $kT_{\rm centre}=5.5\pm0.1$ keV 
is the average gas temperature for the cluster's central 
200$h_{71}^{-1}$ kpc ($2^{\prime}.92$) region. If the conduction 
is suppressed by a factor of $5$ due to the tangled magnetic 
field in the ICM (e.g. Narayan \& Medvedev 2001; Zakamska \& 
Narayan 2003; Chandran \& Maron 2004), the effective conduction time 
increases to $0.3$ Gyr. Consequently, even if the cool gas 
clump was formed through radiative cooling, it would have 
been destroyed very quickly by conduction, which indicates that 
the cool gas clump is most likely caused by merger. 
The same conclusions also hold for off-centre cool gas clumps 
found in other merging clusters (Table 3).

Assuming that the cool gas clump in Abell 3158 has a spherical 
geometry with a diameter of $80-110h_{71}^{-1}$ kpc (Fig. 3$a$), 
and has a constant density, the gas mass is estimated to be about 
$3.0-7.7\times10^{10}$ ${\rm M}_{\odot}$. 
This value is comparable to masses of the off-centre cool  
clumps identified in other merging clusters, which range from about
$2.0\times10^{10}$ ${\rm M}_{\odot}$ 
to 
$1.3\times10^{12}$ ${\rm M}_{\odot}$
(Table 3). It also falls into the gas mass range of the cool-core 
regions of poor clusters and massive groups (e.g. Mulchaey et al. 1996; 
Gastaldello et al. 2007; Baldi et al. 2009). In fact, as shown in Table 3, 
not only the gas mass, but also the size, temperature gradient, and 
the relatively high metal abundance of the cool clump in Abell 3158 
resemble those of other off-centre cool clumps detected in 1E $0657-56$ 
(Markevitch et al. 2002), Abell 754 (Markevitch et al. 2003), Abell 2065 
(Chatzikos et al. 2006), Abell 2255 (Sakelliou \& Ponman 2006), Abell 2256 
(Sun et al. 2002), and Abell 3667 (Vikhlinin et al. 2001). Note that most 
of the off-centre cool clumps in these merging systems are considered to be the 
remnant of the cool-core of either the main cluster or the infalling
sub-cluster. Based on all these similarities, it is natural to speculate that 
the cool gas clump in Abell 3158 was formed by the ram pressure stripping 
or slingshot during a merger event.

The above speculation is enhanced by the study of the correlation 
between gas temperature and entropy, which is defined as 
$S=kT/n_{\rm gas}^{2/3}$ 
in terms of gas temperature $kT$ and density $n_{\rm gas}$. By quoting 
the data of Zhang et al. (2007) and Sanderson, O'Sullivan \& Ponman 
(2009), we re-calculate the redshift-corrected gas entropy at 
$0.1R_{500}$ ($R_{500}$ is the radius at which the mean over-density 
is 500 with respect to the critical density of the universe) for 
a sample of 13 cool-core clusters and a sample of 19 non-cool-core 
clusters. We also calculate the gas entropy of the seven cool gas 
clumps listed in Table 3. The entropy-temperature distribution is 
plotted in Fig. 6, where the strong correlations for the cool-core 
and non-cool-core clusters are characterized by two power-law models, 
respectively (Sanderson et al. 2009). It can be clearly seen that 
cool gas clumps in Abell 3158 and most of other merging clusters 
follow the relation for cool-core clusters, which reveals their 
nature as cool-core remnants. Note that the gas entropy of 
the cool clump in 1E 0657-56 is significantly below the relation 
for cool-core clusters at least 1$\sigma$ limit (Fig. 6), which is possibly 
due to that the cool clump is compressed extremely as it 
experiences the ram pressure stripping force and passes through the deep 
potential of the main cluster during this drastic high-velocity merger event 
(Markevitch et al. 2002).

Based on above data analysis, calculations, and discussions, we 
propose that the appearance of the massive off-centre cool gas clumps
can be regarded as a robust diagnostic of merger state. Combining the 
study of cold fronts (e.g. Owers et al. 2009), which survive longer than 
the shocks, with the study of such cool gas clumps, we may be 
able to place tight constraints on the merger history.

\subsection{Possible merging scenarios}

In Abell 3158, the X-ray bow edge of the faint cold front is pointing 
towards west in projection (\S3.1$-$\S3.3), which indicates that 
the symmetrical axis of the merging process is reasonably close 
to the west-east direction, although the projected effect lowers 
the detectability of the edge. The direction is also roughly consistent 
with the extending direction of the three cD galaxies (Fig. 1$a$), 
as well as the galaxy distribution of the high-velocity subgroup (Fig. 5$b$). 
Straightforwardly, the high-velocity subgroup has already made 
its closest approach to the core of the main cluster roughly 
from east to west. In this scenario, the apparent discrepancy 
between the moving velocity of the cool gas clump 
($700^{+140}_{-340}$ km s$^{-1}$; \S3.3) 
and the relative line-of-sight velocity of the high-velocity subgroup 
(1740$\pm180$ km s$^{-1}$; \S4) 
demonstrates the gas component of the 
high-velocity subgroup possibly has been significantly slowed by 
the ram pressure stripping force. Regarding the late stage of the 
major merger (\S4), perhaps the high-velocity subgroup is in its 
second core passage. On the other hand, because the three cD galaxies 
all lie well behind the cold front (Fig. 1$a$), perhaps the cluster is 
in another merging scenario, in which the cool gas clump has been slingshot
out from the centre of the main cluster, as expected after core passage,
although the optical centroid of the high-velocity subgroup is not behind 
the faint cold front edge.

\section{SUMMARY}
By analysing the {\it Chandra} and {\it XMM-Newton} data of the 
nearby galaxy cluster Abell 3158, we identify  a massive, 
off-centre cool gas clump, which is moving at a 
velocity of $\mathcal M=0.6^{+0.1}_{-0.3}$, toward west on the sky plane 
behind a faint cold front. The possibilities of this substructure 
being formed by inhomogeneous radiative cooling and by central AGN activity
are excluded. Based on the results obtained in the X-ray and optical analysis,  
we speculate that the cool gas clump is the remnant of the original central 
cool-core of the main cluster or the infalling sub-cluster during a major merger event.
This case shows that the appearance of such massive, off-centre cool gas clumps 
can be used to diagnose the dynamical state of a cluster, especially when 
violent merger signatures (e.g. prominent shocks and cold fronts) 
are absent. 

\section*{Acknowledgments}

We thank the {\it Chandra} and {\it XMM-Newton} teams for making this research.
This work was supported by the National Science Foundation of China 
(Grant No. 10673008, 10878001 and 10973010), the Ministry of 
Science and Technology of China (Grant No. 2009CB824900/2009CB24904), 
and the Ministry of Education of China (the NCET Program).


\begin{table*}
\begin{center} 
\begin{tabular}{lcccccccc}
\hline
\hline
\multicolumn{9}{c}{East Sector$^{\it a}$} \\ \cline{1-9}
\hline
Model&Data&$S_{0}$&$\beta$&$r_{\rm c}$&...&...&...&$\chi^2/dof$\\
&&(cnts cm$^{-2}$ s$^{-1}$ arcsec$^{-2}$)&&($h_{71}^{-1}$ kpc)&&&&\\
\hline
single $\beta$&{\it Chandra}&$8.85\pm0.26\times10^{-8}$&0.73$\pm0.02$&197.2$\pm5.4$&...&...&...&28.5/27\\
single $\beta$&    {\it XMM}&$7.94\pm0.07\times10^{-8}$&0.74$\pm0.02$&187.0$\pm1.8$&...&...&...&41.9/27\\
\hline
\hline
\multicolumn{9}{c}{West Sector$^{\it b}$} \\ \cline{1-9}
\hline
Model&Data&$n_{\rm g,1}$&$\beta_{1}(\alpha)$&$R_{\rm c1}(R_{\rm cut})$&$n_{\rm g,2}$&$\beta_{2}$&$R_{\rm c2}$&$\chi^2/dof$\\
&&($10^{-3}$ cm$^{-3}$)&&($h_{71}^{-1}$ kpc)&($10^{-3}$ cm$^{-3}$)&&($h_{71}^{-1}$ kpc)&\\
\hline
A&{\it Chandra}&$3.95\pm0.12$&$4.50\pm0.12$&$243.1\pm4.0$&$2.60\pm0.04$&$0.54\pm0.01$&$257.2\pm2.3$&115.1/15\\
A&{\it XMM}    &$2.90\pm0.06$&$2.88\pm0.06$&$209.1\pm4.0$&$3.08\pm0.04$&$0.50\pm0.01$&$182.3\pm1.2$&72.5/24\\
B&{\it Chandra}&$4.51\pm0.09$&$0.07\pm0.04$&$111.0\pm4.4$&$3.16\pm0.05$&$0.63\pm0.01$&$238.9\pm3.3$&20.7/15\\
B&{\it XMM}    &$4.72\pm0.08$&$0.03\pm0.03$&$104.9\pm2.2$&$3.92\pm0.05$&$0.66\pm0.01$&$200.0\pm2.1$&33.5/24\\
\hline
\end{tabular}
\caption{Best-fits to the exposure-corrected surface brightness profiles (Fig. 2$b$), 
which are extracted in $0.3-2.0$ keV from the 
partial elliptical annuli defined in the east and west sector regions (Fig. $2a$). 
$^{a}$
A single $\beta$ model is sufficient in fitting the SBPs 
extracted in the east sector (\S3.1).
$^{b}$
Best-fit deprojected spectral parameters obtained in \S3.2 (see also Table 2 and Fig. $4e$) 
and two gas density models (\S3.3; model A$=$$\beta$ component+$\beta$ component (Eq. 3);
model B$=$truncated power-law component+truncated $\beta$ component (Eq. 4)) 
are used to fit the {\it Chandra} and {\it XMM-Newton} profiles extracted in the west sector.}
\end{center}
\end{table*}

\begin{center} 
\begin{table*}
\begin{tabular}{lccccc}
\hline
\hline
Region$^{\it a}$&
Radius$^{\it b}$&
Data&
$kT$&
$Z$& 
$\chi^2/dof$  \\
&
($h_{71}^{-1}$ kpc)&
&
(keV) &
(Z$_{\odot}$)& \\
\hline
\multicolumn{6}{c}{PROJECTED$^{\it c}$} \\ 
\cline{1-6}
W1&   0.0$-$45.4&${\it Chandra}$&$5.46^{+0.48}_{-0.31}$&$0.75^{+0.27}_{-0.25}$&119.2/124\\
&               &{\it XMM}      &$5.43^{+0.41}_{-0.30}$&$0.90^{+0.43}_{-0.27}$&124.7/120\\
W2&  45.4$-$95.3&${\it Chandra}$&$4.96^{+0.19}_{-0.19}$&$0.69^{+0.14}_{-0.14}$&199.2/206\\
&               &{\it XMM}      &$4.85^{+0.17}_{-0.17}$&$0.81^{+0.13}_{-0.13}$&192.5/207\\
W3& 95.3$-$136.1&${\it Chandra}$&$5.42^{+0.33}_{-0.23}$&$0.78^{+0.19}_{-0.18}$&229.4/205\\
&               &{\it XMM}      &$5.53^{+0.29}_{-0.23}$&$0.93^{+0.18}_{-0.17}$&166.2/165\\
W4&136.1$-$190.5&${\it Chandra}$&$5.17^{+0.21}_{-0.21}$&$0.69^{+0.15}_{-0.14}$&212.9/206\\
&               &{\it XMM}      &$5.54^{+0.30}_{-0.32}$&$0.48^{+0.14}_{-0.14}$&187.6/187\\
W5&190.5$-$258.5&${\it Chandra}$&$4.86^{+0.21}_{-0.21}$&$0.58^{+0.15}_{-0.14}$&163.5/126\\
&               &{\it XMM}      &$5.16^{+0.19}_{-0.19}$&$0.63^{+0.14}_{-0.13}$&195.5/216\\
\hline
\multicolumn{6}{c}{DEPROJECTED$^{\it d}$}\\ 
\cline{1-6}
W1&   0.0$-$45.4&{\it XMM}&$5.71^{+1.04}_{-0.71}$&$0.99^{+0.87}_{-0.70}$&896.3/893\\
W2&  45.4$-$95.3&{\it XMM}&$3.98^{+0.43}_{-0.41}$&$0.58^{+0.37}_{-0.31}$&...\\
W3& 95.3$-$136.1&{\it XMM}&$5.38^{+0.74}_{-0.51}$&$1.24^{+0.59}_{-0.50}$&...\\
W4&136.1$-$190.5&{\it XMM}&$5.46^{+1.24}_{-1.17}$&$0.02^{+0.53}_{-0.02}$&...\\
\hline
\end{tabular}
\caption{Gas temperature and abundance distributions in the west sector region.
$^{\it a}$
We divide the west sector region into wider partial elliptical annuli, 
i.e., W1$-$W5 regions (Fig. 3$a$).
$^{\it b}$
Equivalent inner and outer radii for each partial elliptical annulus, 
which are defined as $r=\sqrt{ab}$, where a and b are the corresponding
semi-major and semi-minor axes, respectively. 
$^{\it c}$
An absorbed \textsc{apec} model is used to fit the {\it Chandra} ACIS and 
{\it XMM-Newton} MOS1+MOS2+pn spectra extracted in the W1$-$W5 regions.
The redshift and absorption are fixed to $z=0.0597$ and 
the Galactic value $N_{\rm H} = 1.62 \times 10^{20}$ cm$^{-2}$ 
(Dickey \& Lockman 1990), respectively. Errors are quoted at the 68\% confidence. 
$^{\it d}$
The \textsc{projct} model coded in \textsc{xspec} v12.4.0 is used to 
deproject the {\it XMM-Newton} MOS1+MOS2+pn spectra.
Errors are quoted at the 90\% confidence.}
\end{table*}
\end{center}

\begin{center} 
\begin{table*}
\begin{tabular}{lcccccccc}
\hline
\hline
Host cluster&
&
1E 0657-56&
A754&
A2065&
A2255&
A2256&
A3158&
A3667\\
\hline
Clump's scale $l$&($h_{71}^{-1}$ kpc)&70&120&60&100&160&$80-110$&300\\
Projected distance to centre $d$&($h_{71}^{-1}$ kpc)&650&200&60&130&130&85&400\\
Gas density $n_{\rm gas}$&($10^{-3}{\rm cm}^{-3}$)&24&4.6&7.2&$2.6-4.3$&4.3&4.5&3.8\\
Gas temperature $T$&(keV) &$7.0\pm1.0$&$6.2\pm0.1$&$3.4\pm0.6$&$5.2\pm0.7$&$4.5\pm1.0$&$4.0\pm0.4$&$4.1\pm0.2$\\
Ambient temperature $T_{\rm amb}$& (keV)&13.6&7.4&5.5&6.9&6.7&5.5&7.7\\
Cluster temperature $T_{\rm cluster}$& (keV)&$13.6\pm1.0$&$10.0\pm0.3$&$5.5\pm0.2$&$6.9\pm0.3$&$6.7\pm0.2$&$5.5\pm0.1$&$7.3\pm0.2$\\
Metal abundance $Z$ &(Z$_{\odot}$)&...&$0.54$&$1.36$&...&$0.84$&$0.81$&$0.81$\\
Cooling time $t_{\rm cool}$ &($10^{9}$ yr)&2.4&12.0&5.7&$11.7-19.4$&10.9&9.8&11.8\\
Effective conduction time $t_{\rm cond}$&($10^{9}$ yr)&0.1&0.3&0.3&$0.2-0.3$&0.7&$0.3-0.6$&1.5\\
Gas mass $M_{gas}$ &($10^{11}{\rm M}_{\odot}$)&1.1&1.0&0.2&$0.3-0.6$&2.3&$0.3-0.8$&13.4\\
Gas entropy $S$&(keV cm$^{2}$)&$84\pm12$&$224\pm4$&$91\pm16$&$236\pm80$&$170\pm38$&$147\pm15$&$168\pm8$\\
\hline\\
\end{tabular}
\caption{Properties of the massive, off-centre cool gas clumps,
which are detected in  
1E 0657-56 (Markevitch et al. 2002),
Abell 754 (Markevitch et al. 2003; Henry et al. 2004), 
 Abell 2065 (Chatzikos et al. 2006), Abell 2255 (Sakelliou \& Ponman 2006), 
Abell 2256 (Sun et al. 2002), and Abell 3667 (Vikhlinin et al. 2001; 
Briel et al. 2004), respectively. Cooling times of the cool clumps are 
calculated from Eq (5), and effective conduction times are esitmated 
from Eq (6) by assuming a tangled magnetic effect.   
Gas masses of the cool clumps are calculated by assuming that
the gas density, temperature, and metal abundance are constant in the clump.
Gas entropies are given by $S=kT/n_{\rm gas}^{2/3}$.}
\end{table*}
\end{center}
 
\clearpage

\begin{figure*}
\begin{center}
\includegraphics[width=18cm,angle=0]{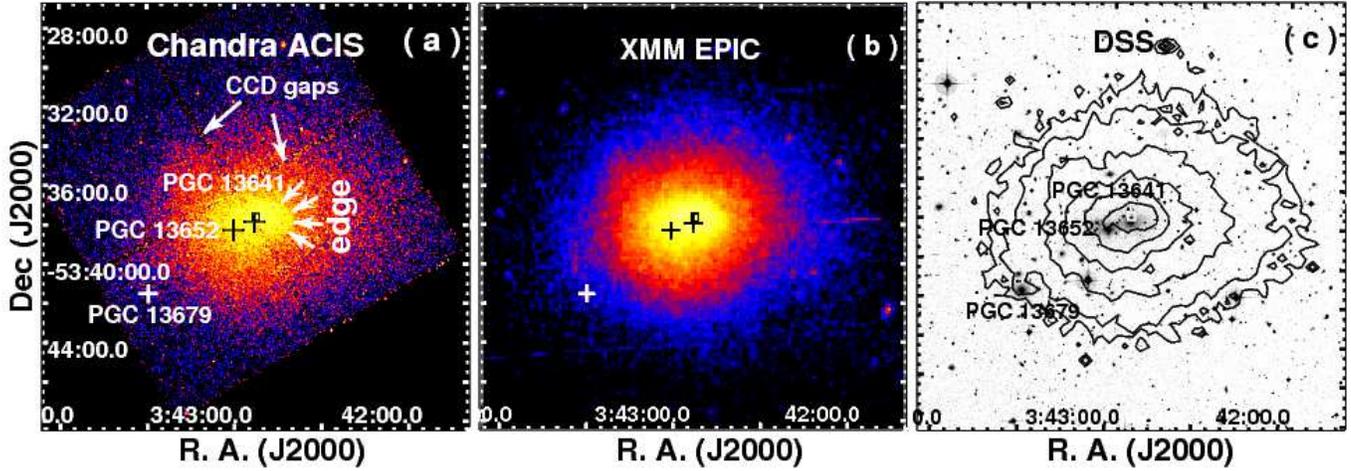}
\caption{
($a$)--($b$) Adaptively smoothed and exposure-corrected {\it Chandra}
ACIS-I (ObsID 3712) and {\it XMM-Newton} EPIC (ObsID 0300210201)
images of the central 1.5$h_{71}^{-1}$ Mpc$\times$1.5$h_{71}^{-1}$ Mpc
($21^{\prime}.93\times21^{\prime}.93$) region of Abell 3158. Both images 
are extracted in 0.3$-$2.0 keV and plotted in logarithmic scale. 
One small box and three crosses are used to mark the peak
of the X-ray halo and the centres of luminous member galaxies PGC 13641, PGC 13652, and PGC 13679, 
respectively. ($c$) Optical DSS {\it B}-band image for the same sky field 
as ($a$) and ($b$), on which the {\it Chandra} X-ray intensity contours 
(square root scale) are plotted. }
\end{center}
\end{figure*}

\begin{figure*}
\begin{center}
\includegraphics[width=18cm,angle=0]{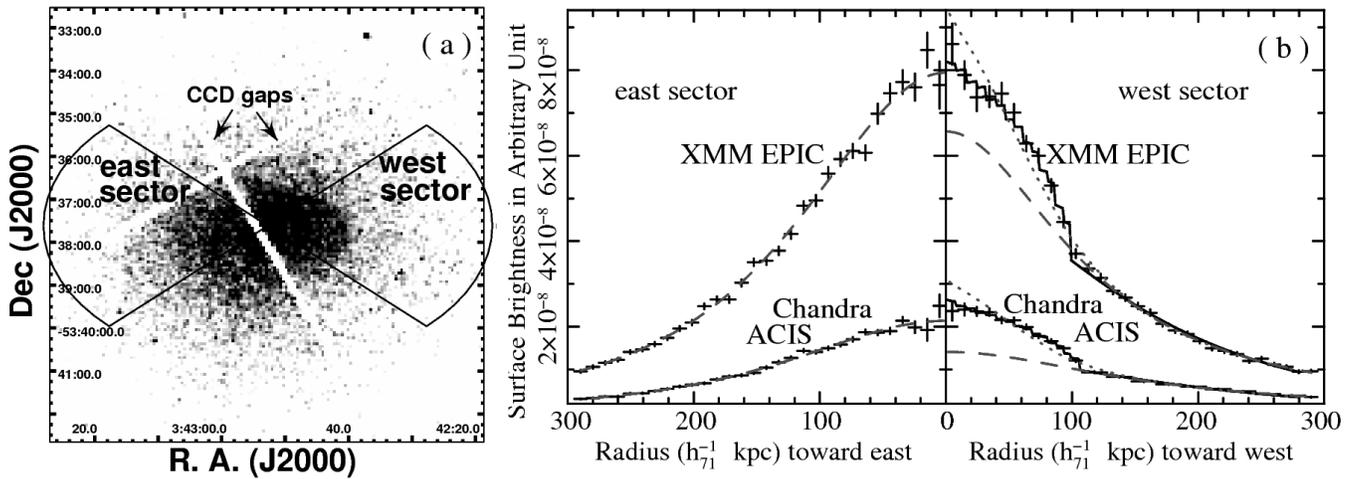}
\caption{ ($a$) 
East and west sector regions used to extract the X-ray SBPs, which are shown in ($b$). 
Here the {\it Chandra} ACIS image is used as the background image.
($b$) {\it Chandra} ACIS and {\it XMM-Newton} EPIC 0.3$-$2.0 keV SBPs 
extracted from the partial elliptical annuli defined in the two sector regions.
For SBPs extracted in the east sector, a single $\beta$ model 
can give an acceptable fit to the observed SBPs (dashed lines; \S3.1 and Table 1). 
For SBPs extracted in the west sector, dashed, dotted, and solid lines 
are used to show the best-fits obtained with the single $\beta$ 
model (\S3.1), two-$\beta$ model, and truncated powerlaw+$\beta$ model 
(\S3.3 and Table 1), respectively.  
}
\end{center}
\end{figure*}

\begin{figure*}
\begin{center}
\includegraphics[width=16cm,angle=0]{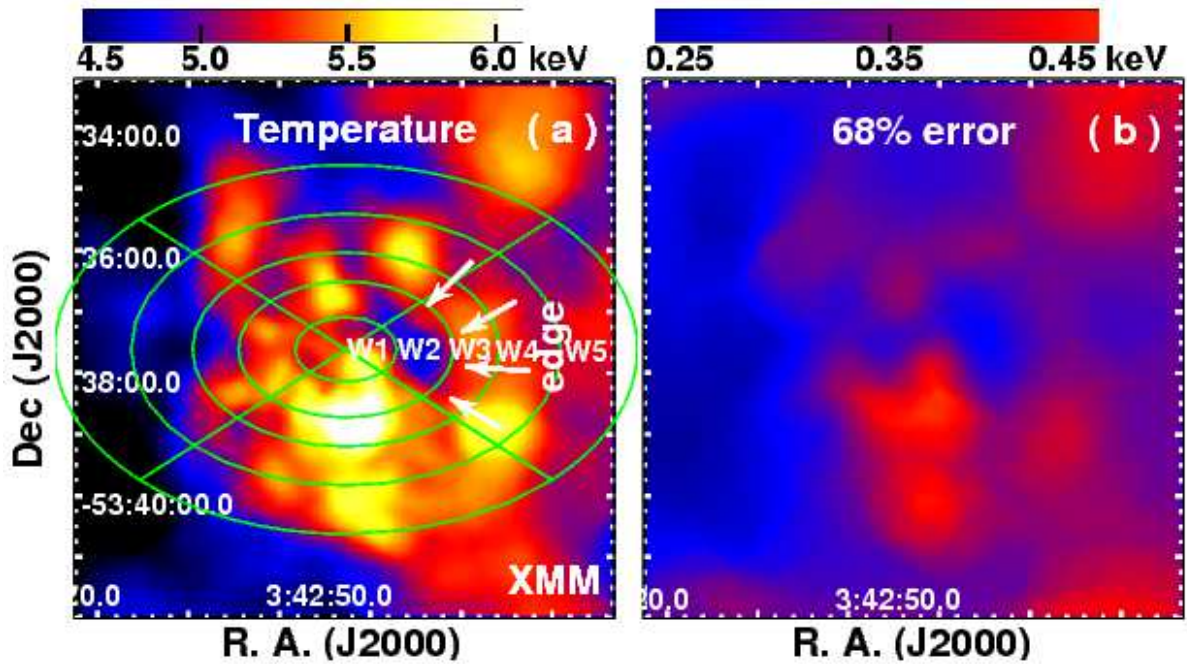}
\caption{ 
($a$) Projected gas temperature map for the central 
0.6$h_{71}^{-1}$ Mpc$\times$0.6$h_{71}^{-1}$ Mpc ($8^{\prime}.77\times8^{\prime}.77$) region,
along with the 68\% error map ($b$), which are calculated from the 
{\it XMM-Newton} MOS1+MOS2+pn data. Partial elliptical annuli 
defined in the north, east, south, and west sector regions 
are shown in ($a$), which are used to extract both the {\it Chandra} 
and {\it XMM-Newton} spectra to be analysed in \S3.2. 
}
\end{center}
\end{figure*}

\begin{figure*}
\begin{center}
\includegraphics[width=16cm,angle=0]{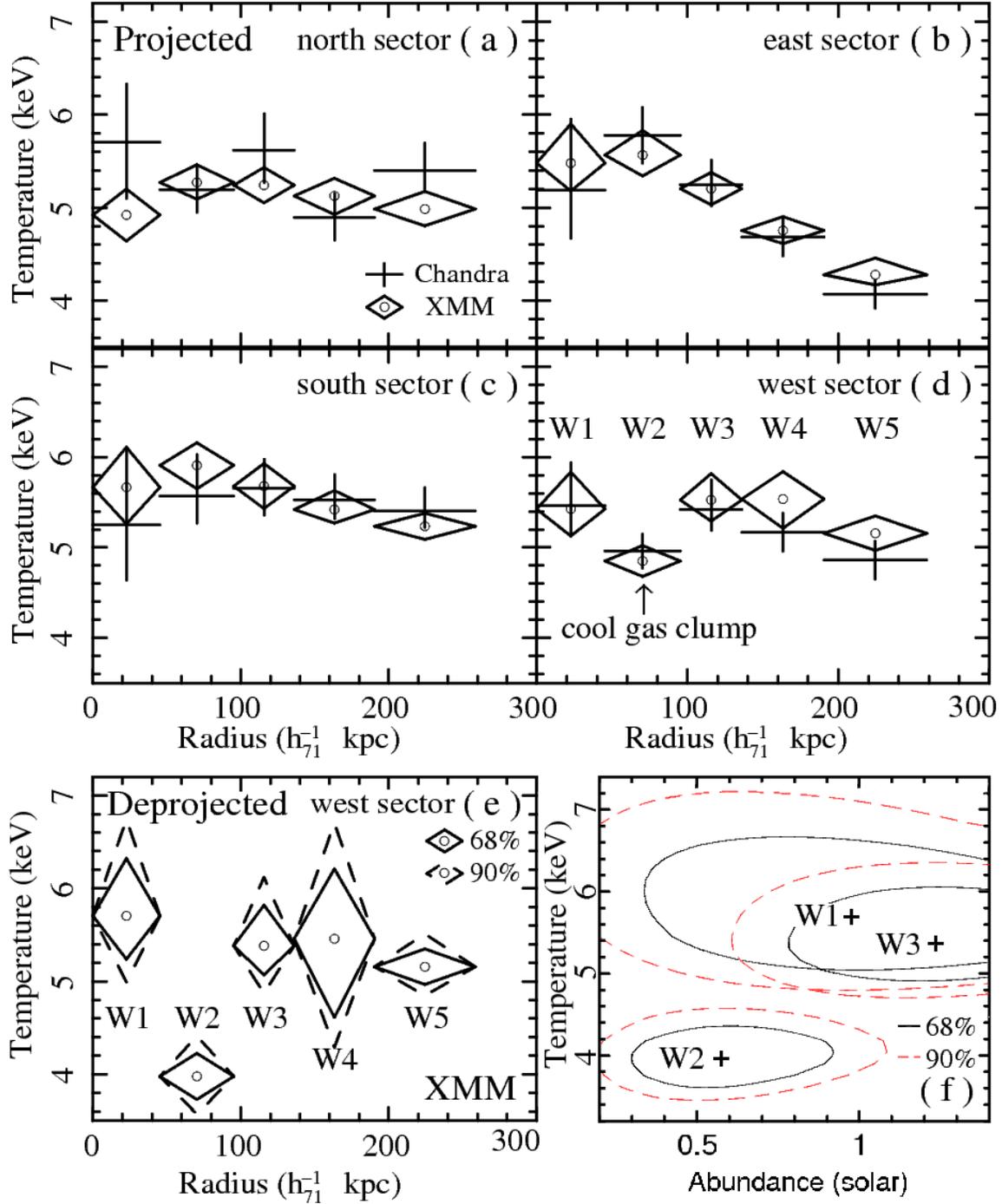}
\caption{ 
($a$)$-$($d$) Projected gas temperature profiles obtained with 
{\it Chandra} ACIS (cross) and {\it XMM-Newton} MOS1+MOS2+pn (diamond)
data for the north, east, south, and west sectors defined in Fig. 3$a$, 
along with the 68\% errors. 
($e$) Deprojected {\it XMM-Newton} temperature profiles for the west sector,
along with the 68\% (solid) and 90\% (dashed) errors (see also Table 2). 
($f$) Fit-statistic contours of temperature and abundance at the 68\% 
($\Delta\chi^{2}=2.30$; solid) and 90\% ($\Delta\chi^{2}=4.61$; dashed) 
confidence levels for the inner three partial annuli of the west sector, 
all obtained in the {\it XMM-Newton} deprojected analysis.  
}
\end{center}
\end{figure*}

\begin{figure*}
\begin{center}
\includegraphics[width=16cm,angle=0]{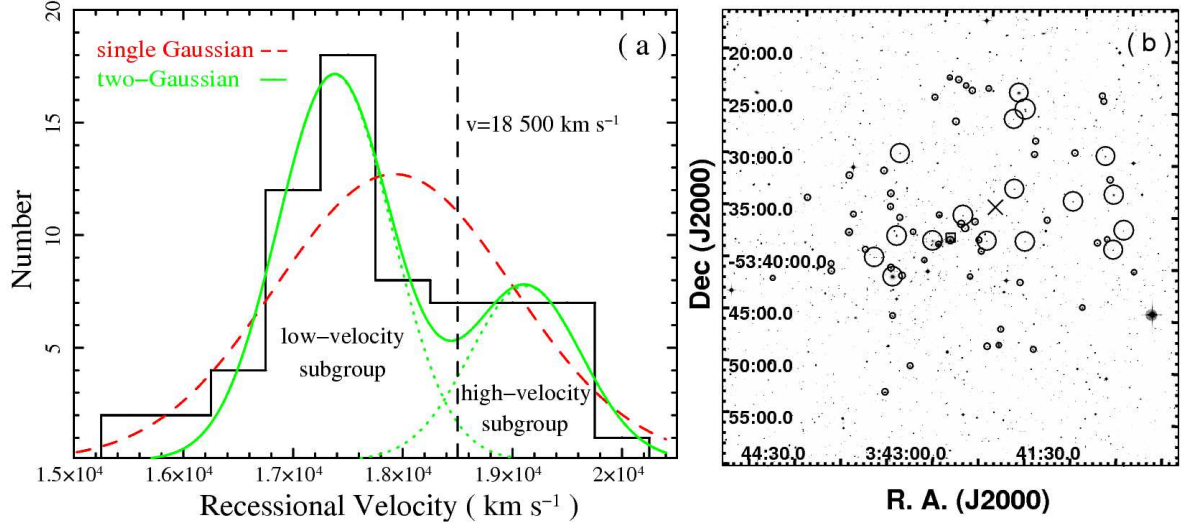}
\caption{
($a$) Line-of-sight velocity distribution of 68 member galaxies identified
in Abell 3158. Original velocity data are drawn from Smith et al. (2004). 
The distribution is fitted with a two-Gaussian model (solid) and 
a single Gaussian model (dashed; \S4), respectively. 
($b$) DSS optical image for the central 
3.0$h_{71}^{-1}$ Mpc$\times$3.0$h_{71}^{-1}$ Mpc ($43^{\prime}.86\times43^{\prime}.86$) 
region of Abell 3158, where the low-velocity ($v<18~500$ km s$^{-1}$) 
and high-velocity ($v>18~500$ km s$^{-1}$) member galaxies 
are marked with small and large circles, respectively. 
The X-ray peak is marked with a small box and the geometric centroid of  
the high-velocity subgroup is marked with a cross.  
}
\end{center}
\end{figure*}

\begin{figure*}
\begin{center}
\includegraphics[width=12cm,angle=0]{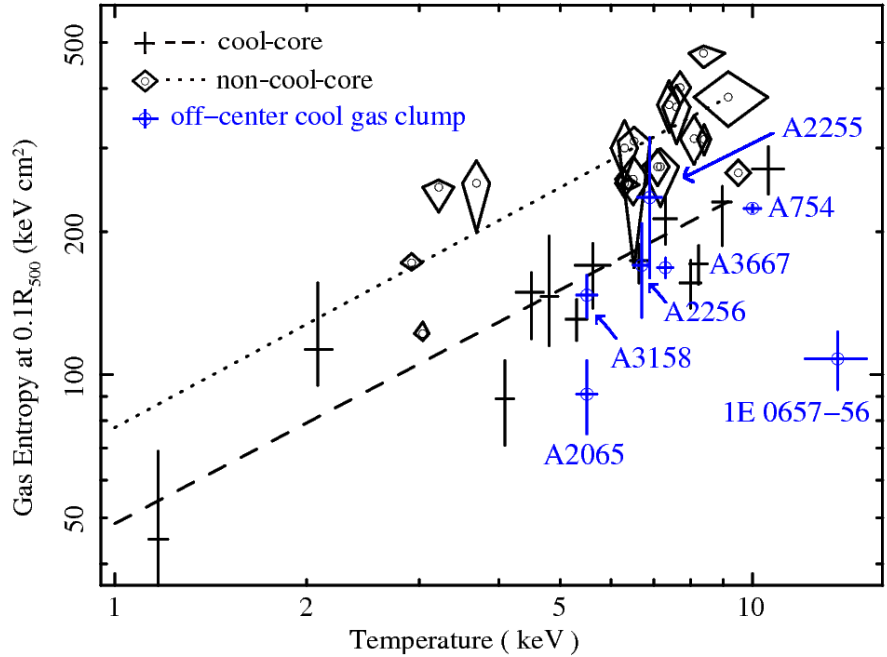}
\caption{ 
Redshift-corrected gas entropy measured at $0.1R_{500}$ (roughly $50-150$ kpc) vs average gas temperature 
for cool-core clusters (crosses) and non-cool-core clusters (diamonds), 
which are re-calculated based on the data of Zhang et al. (2007) 
and Sanderson et al. (2009).
The corresponding best-fit power law models 
given by Sanderson et al. (2009) are also shown as dashed and dotted lines, respectively.  
The entropy-temperature distribution of the off-centre cool gas clumps 
(Table 3) is also plotted.}
\end{center}
\end{figure*}

\label{lastpage}

\end{document}